# Dual-comb spectroscopy for high-temperature reaction kinetics

NICOLAS H. PINKOWSKI,[1*] YIMING DING,[1] CHRISTOPHER L. STRAND,[1*] AND RONALD K. HANSON[1]

[1]High Temperature Gasdynamics Laboratory, Department of Mechanical Engineering, Stanford University

RAPHAEL HORVATH,[2] MARKUS GEISER,[2]

[2] IRsweep AG, Laubisruetistr. 44, 8712 Staefa, Switzerland

*Corresponding author: npinkows@stanford.edu

**Abstract**

In the current study, a quantum-cascade-laser-based dual-comb spectrometer (DCS) was used to paint a detailed picture of a 1.0 ms high-temperature reaction between propyne and oxygen. The DCS interfaced with a shock tube to provide pre-ignition conditions of 1225 K, 2.8 atm, and 2% p-$C_3H_4$/18% $O_2$/Ar. The spectrometer consisted of two free-running, non-stabilized frequency combs each emitting at 179 wavelengths between 1174 and 1233 cm$^{-1}$. A free spectral range, $f_r$, of 9.86 GHz and a difference in comb spacing, $\Delta f_r$, of 5 MHz, enabled a theoretical time resolution of 0.2 µs but the data was time-integrated to 4 µs to improve SNR. The accuracy of the spectrometer was monitored using a suite of independent laser diagnostics and good agreement observed. Key challenges remain in the fitting of available high-temperature spectroscopic models to the observed spectra of a post-ignition environment.

## 1. Introduction

Energy systems involving high temperatures and fast reactions rely on absorption spectroscopy for fundamental research and field diagnostics. However, measurements at extreme conditions are fundamentally challenging due to short timescales and complex thermochemistry that is often present. For such environments, high-bandwidth measurement of a spectral surface (absorbance as a function of wavenumber and time) could provide sufficient information for speciation, thermometry, and knowledge of quantum state populations in non-equilibrium environments. However, many broadband measurement techniques, such as Fourier Transform Infrared spectrometers or rapid-tuning broad-scan external cavity quantum cascade lasers [1], are mechanically limited to low time-resolutions. Advanced broadband diagnostics being developed without such mechanical limitations involve super-continuum and frequency-comb absorption spectroscopy [2–9].

Two frequency combs can be combined to help interpret the signal of multiple wavelengths simultaneously, a technique called dual-comb spectroscopy (DCS). Notable studies involving DCS include: the first dual-comb field diagnostic monitoring greenhouse gases [2], dual-combs in a rapid compression machine [8], and DCS used to study protein dynamics [10].

It is of great interest to apply DCS to study high-temperature reaction kinetics. However, this application puts stringent constraints on a spectrometer's capabilities. At high-temperatures, reactions occur over short timescales and often in vessels with short path lengths. Correspondingly, it is important to have microsecond time-resolution and a high power across all wavelengths (power-per-wavelength) to ensure sufficient signal to noise in a dynamic environment with emission and vibration. Furthermore, it is desirable for measurements to be made in the mid-infrared between 3-20 µm with access to the prominent absorption features of hydrocarbons, such as $CH_2$ and $CH_3$ stretching modes in the 3 µm region, –CH=$CH_2$ bending modes near 11 µm, or CH bending modes near 15 μm [11]. To the authors' knowledge, reaction kinetics using frequency-comb spectroscopy was first studied by Fleisher et al. & Bjork et al. involving measurements of a photolysis-induced reaction at 25 μs time resolution [12,13]. These studies did not use DCS but used a single frequency comb and diffraction grating to interpret the signal. The use of two frequency combs, dual-comb spectroscopy, has been applied to high-temperature, non-reacting conditions in a rapid compression machine by Draper et al. at 704 μs time-resolution and also by Schroeder et al. in the exhaust of a gas turbine at high temperatures and 10-60 s time-resolution [4,8]. Draper et al. and Schroeder et al. achieved exceptional spectral resolution with tens of thousands of wavelengths in the near-IR. A worthy extension of the work of Draper and Schroeder et al. would be to establish similar techniques in the mid-IR with improved time-resolution.

Efforts to extend DCS into the 3-20 μm region have involved distributed feedback generation DFG [14], optical parametric oscillators (OPO) [15–17], and quantum-cascade-laser (QCL) approaches [6,7,18–20]. Obtaining coherency between two frequency combs is an enabling factor for DCS and DFG approaches can achieve this readily. However, DFG methods generally have limited power-per-wavelength, which limits application in many environments [19]. Alternatively, OPO-based combs can obtain a higher power-per-wavelength while also providing access to the mid-infrared. However, a disadvantage of OPO combs involves the fact that coherency between separate combs is more challenging to obtain. For this reason, dispersion was used when Fleisher et al. and Bjork et al. first studied reaction kinetics with a mid-infrared OPO comb [12,13]. An extension of the work of Fleisher and Bjork et al. would involve the use of two coherent, mid-IR frequency combs configured as a dual-comb spectrometer.

Prominently, a class of OPO combs operating in degeneracy, known also as divide-by-two subharmonic OPOs, have successfully demonstrated coherency and dual-comb operation [17,21,22]. The OPO-based DCS of Muraviev et al. [17] offered remarkable spectral coverage (3.1-5.5 μm with 350,000 spectral data points), although with a minimum time-resolution of only 7 ms. While only a matter of time, OPO-based DCS has yet to demonstrate both dual-comb and microsecond operation in the mid-infrared.

Advances in the field of quantum-cascade-laser (QCL) frequency combs have led to the development of mid-IR, solid-state spectrometers that have attracted considerable attention in the recent literature [7,10,20,23–25]. In particular, the recent advent of QCL frequency combs provides an alternative to DFG and OPO combs when a high power-per-wavelength and fast time resolution are needed [20,25]. QCL combs can offer over 1 mW per wavelength and microsecond time-resolution, with the tradeoff of having a larger comb spacing (~0.3 cm$^{-1}$) and less spectral range than alternative systems. To date, these QCL-based systems have performed well in proof-of-concept demonstrations, e.g. to observe protein dynamics [10,20].

In this study, we demonstrate a mid-infrared QCL-based dual-comb spectrometer (DCS) capable of microsecond-resolved measurements of energetic gas phase reactions. Specifically, an aggressive chemical reaction, propyne oxidation, was studied using a pressure-driven shock tube equipped with two QCL frequency combs and a suite of independent validation diagnostics. This study details the dual-comb spectrometer employed, provides a demonstrative propyne oxidation measurement, and concludes with an assessment of system accuracy.

## 2. Background

### 2.1. Laser systems

An overview of the experimental setup is provided in Fig. 1, illustrating the placement of frequency combs and other validation diagnostics on a shock tube with a bore of 14.13 cm. The shock tube was used to generate the high-temperature conditions and was described by Strand et al. [1] previously. The shock tube interfaced with a dual-comb (QCL-based) spectrometer, and two additional interband cascade lasers (ICL) to study water and propyne. A side-wall pressure transducer was also used to monitor the pressure during the reaction.

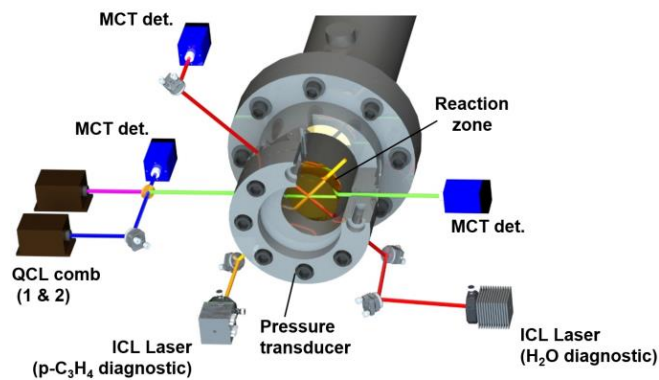

**Fig. 1.** An illustration of a shock tube equipped with a QCL dual-comb spectrometer and supporting validation diagnostics. HgCdTe (MCT) detectors, interband cascade lasers (ICLs), pressure transducer, and a reaction zone are indicated.

*2.1.1. Dual-comb spectrometer*

The DCS developed for the current study relies on two free-running QCL-based frequency combs (from IRsweep AG), each spanning about 60 cm$^{-1}$. The native emission power of each comb was >500 mW resulting in >2 mW per comb tooth on average. Each QCL comb was slightly detuned from another to yield two different repetition frequencies, $f_1$ and $f_2$ that were both at approximately 0.3 cm$^{-1}$ or 9.86 GHz. The difference in repetition rates, $\Delta f_{rep} = f_1 - f_2$, was typically around 5 MHz. This slight difference resulted in the creation of a set of beating frequencies between each detuned pair of spectral lines. This set of beat-notes constitutes an additional frequency comb in the radio-frequency (RF) domain. Unlike the mid-infrared combs that generated the beat-notes, this RF comb is at much slower frequencies and its intensity vs. time can be measured on a detector.

The combs emitted at 179 different wavelengths from 1174 to 1233 cm$^{-1}$ providing access to high-temperature absorption features for alkynes and water in this region [26,27]. FTIR measurements of the corresponding emission from each comb are presented in Fig. 2. The narrow emission of each comb tooth is below the 0.3 cm$^{-1}$ spectral resolution of the FTIR and this under-sampling manifests itself as the broadened peaks seen in Fig 2. However, the FTIR measurements in Fig. 2 very effectively illustrate the emission range of the QCL's and the spectral power distribution that can be expected from the spectrometer.

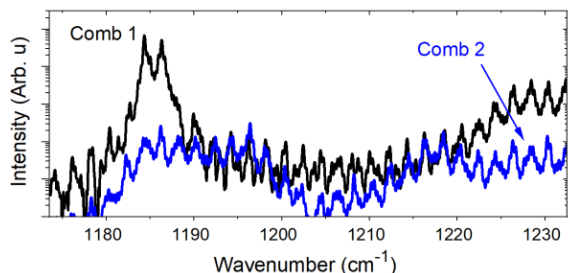

**Fig. 2.** FTIR measurements of each frequency comb. While emission of each comb line is well below the spectral resolution of the FTIR (0.3 cm$^{-1}$), these FTIR measurements provide a clear picture of the wavenumber-dependent power distribution of the system with high power near 1185 cm$^{-1}$ and 1220 cm$^{-1}$ and low power near 1205 cm$^{-1}$.

The emission from the pair of QCL combs was coaligned and split to yield two combined beams. The first (reference) beam was attenuated and focused onto a reference HgCdTe (MCT) detector, while the second (sample) beam was attenuated and transmitted through wedged, anti-reflection-coated ZnSe windows in the shock tube. After traversing the shock tube, this beam was directed through a long-pass filter and focused on another MCT detector. The superimposed wavelengths of light from the two frequency combs heterodyned on each detector and generated a comb in the RF domain spaced at $\Delta f_{rep}$. The detectors had a bandwidth of 1 GHz and were digitally sampled at 2.0 GS/s to record a sequence of interferograms spaced at a period of $1/\Delta f_{rep}$ (0.2 µs). The shortest achievable time resolution of this configuration is determined by the period of a single interferogram and was therefore 0.2 µs. However, the interferograms were then averaged over a time-interval of 4 µs to improve signal-to-noise ratio. In the time domain, a representative signal from the DCS is shown in Fig. 3, which demonstrates a typical interferogram pattern observed.

Figure 4 presents the result of a Fourier transformation of the time-domain data to assess the frequency content of the DCS signal. As can be seen in Fig. 4, the slight detuning between the two frequency combs manifests itself as an additional frequency comb in the RF domain. The beating, or heterodyning, between DCS comb teeth creates the multi-heterodyne pattern in Fig. 4. The discrete and equally spaced peaks confirm that the system is generating a successful multi-heterodyne signal and provides insights about the QCL combs that generated the pattern. Prominently, the RF frequency comb in Fig. 4 has a similar shape as the power profiles from the FTIR measurements presented in Fig. 2.

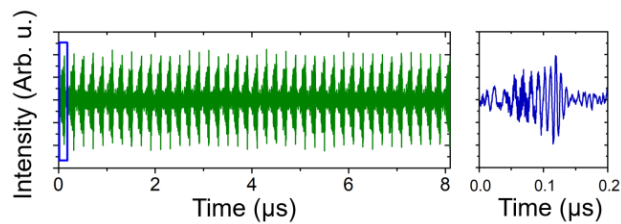

**Fig. 3.** A representative DCS signal in the time domain for multiple periods (left) and for a single period (shown right) with a shared ordinate axis. Good signal is observed for each period and SNR is improved further through time-integration to 4 µs.

The power in each RF beat-note is proportional to the product of the electric fields of the two contributing laser modes. Consequently, changes to the intensities of the measured beat-notes directly correlates to attenuation of light in the optical domain. To calculate absorbance using the DCS system, the attenuation at each optical wavelength was calculated, adjusted by a pre-experimental background measurement, and normalized by an unattenuated signal. The absorbance of the mixture at each wavelength was then determined using the Beer-Lambert relation.

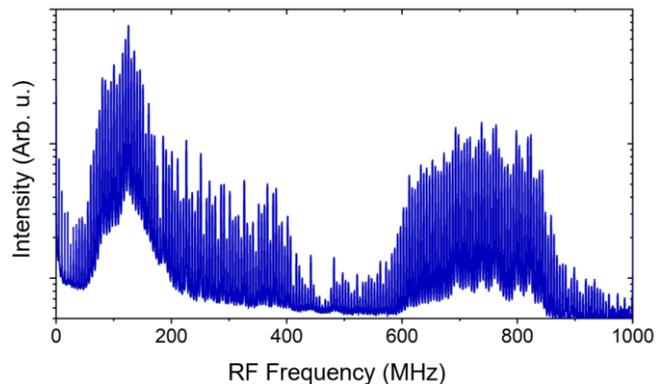

**Fig. 4.** A representative RF, multi-heterodyne pattern of the dual-comb system generated by taking the Fourier transformer of time-domain data. The sharp set of beat-notes presented here indicates effective detuning and operation of the of two QCL combs.

Notably, the spectral separation of the superimposed laser lines contributing to an individual beat-note varied from 50 to 900 MHz; however, an average frequency for each superimposed pair of comb teeth was used for absorbance. Accordingly, the optical resolution of each emission line was limited to 900 MHz (0.03 cm-1), and overall spectroscopic sampling resolution governed by the comb spacing of 0.3 cm$^{-1}$. To ensure quality measurements, the system was calibrated before and after each experiment against the well-known spectrum of polyethylene. Ultimately, while no active wavelength stabilization method was used during an experiment, a comparison of the pre- and post-shock calibration indicated drift of < 100 kHz.

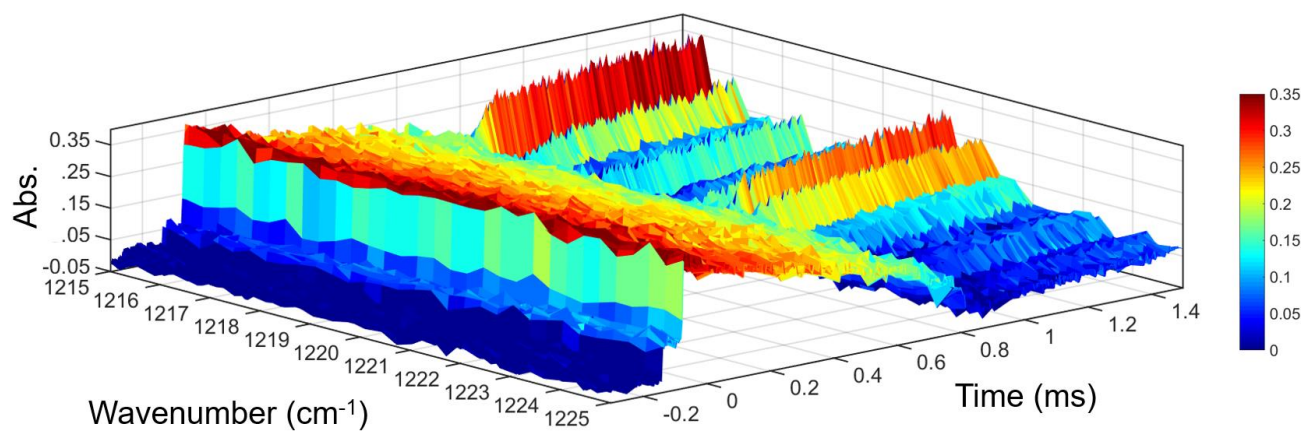

**Fig. 5.** DCS data from 1215 cm$^{-1}$ to 1225 cm$^{-1}$ of propyne oxidation (2% p-C$_3$H$_4$ T$_0$ = 1225 K, P$_0$ = 2.8 atm) illustrating the arrival of the incident and reflected shock before time-zero. The broadband absorption feature of propyne is visible from 0 to 0.6 ms and can be seen transforming into a finely featured spectrum (water) after 0.8 ms. The DCS data demonstrates good SNR during the passage of each shock wave, occurrence of a combustion reaction, and temperature/pressure increase of roughly 2500 K/60x over only 1 ms.

*2.1.2. Supporting laser diagnostics*

In addition to the DCS, two separate laser diagnostic systems, shown in Fig. 1, were employed simultaneously to monitor the reaction. A scanned-wavelength diagnostic for water was used to measure a water absorption transition at 4029.5 cm$^{-1}$ using a wavelength-tuned interband cascade laser (ICL) with a scan repetition rate of 5 kHz [28,29]. Secondly, a fixed-wavelength laser diagnostic for propyne detection was used at 2999.55 cm$^{-1}$ using an additional ICL. Each ICL was coupled with a HgCdTe (MCT) detector that was sampled at 5 MHz.

**2.2. Propyne Oxidation**

Propyne oxidation was selected as the reaction of interest due to a blended broadband absorption feature of propyne and wealth of narrow absorption transitions of water between 1174 -1233 cm$^{-1}$. The absorption feature of propyne is derived from the P-branch of the second harmonic of propyne's –CH in-plane vibrational bending mode [11,26]. At high temperatures, nearly the entire P-branch of the propyne absorption feature is accessible using the DCS system. The absorption features of water are derived from the P-branch of an in-plane symmetric bending mode [11]. During oxidation, the mixture's spectrum evolved in time as the broadband spectrum of propyne transformed into the finely featured spectrum of water.

## 3. Experimental Results

**3.1. Overview**

Data from all diagnostics are presented in Fig. 5-7 for an example shock tube experiment. The pressure-driven, stainless steel shock tube shown in Fig. 1 was used to propagate a shock wave at Mach 2.2 into a test gas mixture with an initial molar composition of 2% p-C$_3$H$_4$/18% O$_2$/80% Ar, a temperature of 298 K, and a pressure of 75 Torr. Figure 5 presents a sample of the DCS data from this reaction where the broadband propyne absorption feature can be seen at early times and the finely featured spectrum of water observed after 0.75 ms. The full dataset of the DCS absorbance measurements is presented in Fig. 6(b). The lowest pressure and absorbance measurements in Fig. 5 and Fig. 6(a-b), located before -75 μs, indicate this pre-shock region. Upon the arrival of the first shock wave, there is a step change in pressure and absorbance of the test gas, seen in Fig. 5 and Fig. 6(a-b) at -75 μs, as the mixture is heated and pressurized to 720 K and 0.7 atm. At this time, the tail of the propyne absorption feature first appears in Fig. 5 and Fig. 6(b) near 1225-1230 cm$^{-1}$ and the apex of the feature rests just beyond the domain limit of 1233 cm$^{-1}$.

After 75 μs, the reflected shock wave arrives (defining time-zero) and provides another pressure increase that is visible in Fig. 6(a). At time-zero, chemistry is assumed to be momentarily frozen and the mixture condition changed to 1225 K and 2.8 atm. Also at time-zero, the system undergoes a rapid redistribution of the state populations, which shifts the crest of the propyne feature seen in Fig. 6(b) to approximately 1220 cm$^{-1}$. Soon after, oxidative pyrolysis begins and propyne is slowly consumed for the subsequent 0.75 ms. During the pyrolysis process, propyne was observed on the separate propyne-specific diagnostic and showed agreement to the overall decomposition rate registered by the DCS, shown on the left of Fig. 7.

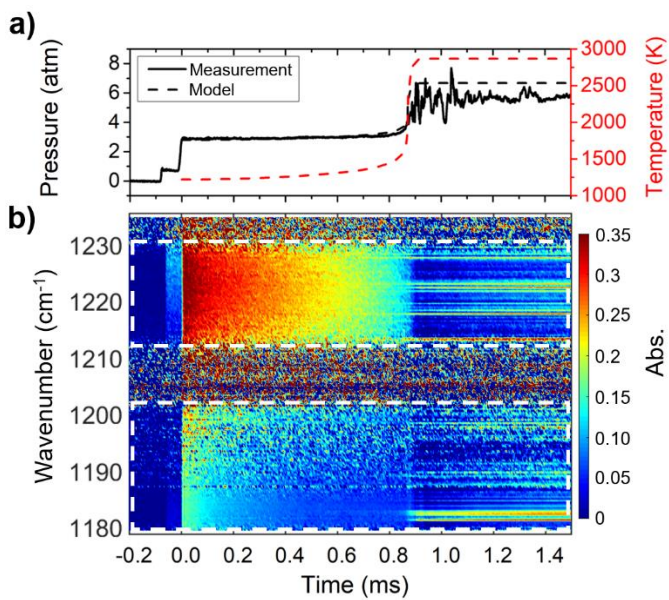

**Fig. 6.** DCS measurements of propyne oxidation with time-zero conditions of 2% p-$C_3H_4$/18% $O_2$ in Ar, 1225 K, and 2.8 atm. All data are aligned in time and share the same x-axis. (a) Measured and simulated thermodynamic conditions for the reaction. (b) Dual-comb-measured absorbance spectrum evolving in time at 4 μs time resolution (See Visualization 1). Two regions with high SNR are delineated in dashed white areas. All simulations shown in (a) used the USC Mech II kinetic mechanism and assumed constant volume [30].

After a residence time of approximately 0.75-0.9 ms, exponential growth of the radical pool takes hold and the mixture ignites. Formation of water during ignition was observed by both the DCS and the scanned-wavelength water diagnostic. In Fig. 6(b), water formation is represented by a brush of fine absorption lines just before 0.9 ms. These absorption features arise coincident with $H_2O$ formation observed using the ancillary water diagnostic, which is shown on the right of Fig. 7.

The current study also includes a video in order to demonstrate the time evolution of the measured spectrum, pressure, and water mole fraction. This video is attached in the supplemental material as Visualization 1.

### 3.2. Speciation

Quantitative speciation measurements were performed using the DCS and the supporting laser diagnostics and compared to a kinetic model (USC Mech II [30]), all shown in Fig. 7. For DCS measurement of propyne and water mole fractions, a vectorized Beer-Lambert system was formulated as presented in Eq. 1. The system included: $W$ a weighting parameter matrix, $K$ a cross-section matrix (cm$^{-1}$ atm$^{-1}$), $x$ a mole fraction vector, and $b$ (cm$^{-1}$ atm$^{-1}$) a vector of the pressure- and path-length-normalized absorbance at each DCS wavelength as a function of time. The formulation of the Beer-Lambert system is discussed in detail in supporting literature [31].

$$\text{Minimize:} \quad \|W(Kx - b)\|_2^2 \quad \textbf{Eq. 1}$$

The diagonal of the weighting matrix was set as the wavenumber- and time-dependent SNR of the DCS system (see Fig. A.1). The SNR of each DCS wavenumber was defined as the absorbance at a given time divided by a measurement of the standard deviation at each wavenumber. The standard deviation at each DCS wavenumber was calculated from a statistical analysis of the DCS noise, which was determined by measuring the absorbance of a non-absorbing medium. Propyne and water were the only species included in the cross-section matrix, $K$. Expected intermediate species that form in trace amounts, but potentially detectable quantities, include: allene, acetylene, methane, ethylene, and formaldehyde. Of these potentially interfering species, only methane and acetylene absorb in this region, although acetylene absorbs very weakly.

Temperature-dependent absorption cross-sections were determined for propyne through a series of additional non-oxidative 2% p-$C_3H_4$/98% Ar experiments ranging from 1100 to 1400 K. Water cross-sections were collected from a series of fuel-lean hydrogen oxidation experiments (2-4%$H_2$/2-4% $O_2$ at 1670-2030 K) and methane-oxidation experiments (4.5% $CH_4$/9% $O_2$/Ar near 2900 K) to generate water in the post-ignition environments[30]. These independent experiments provided the absorption cross-sections that were not available through existing experimental or theoretical databases. Cross-section correlations across all 179 wavelengths were assumed to have a linear temperature dependence. Acetylene and methane cross-section measurements were also collected due to their potential to absorb and interfere in this spectral region. From the additional cross-section measurements, acetylene was undetectable, but methane's cross-section was non-negligible. Equation 1 was solved both with and without methane in order to assess whether it was detectable as an intermediate. Including methane in the absorption cross-section matrix did not yield statistically significant methane measurements but did contribute approximately 5% additional uncertainty to the propyne mole fraction measurements between 0.5 and 1 ms. This additional uncertainty was accounted for in the 95% confidence interval included with the data in Fig. 7.

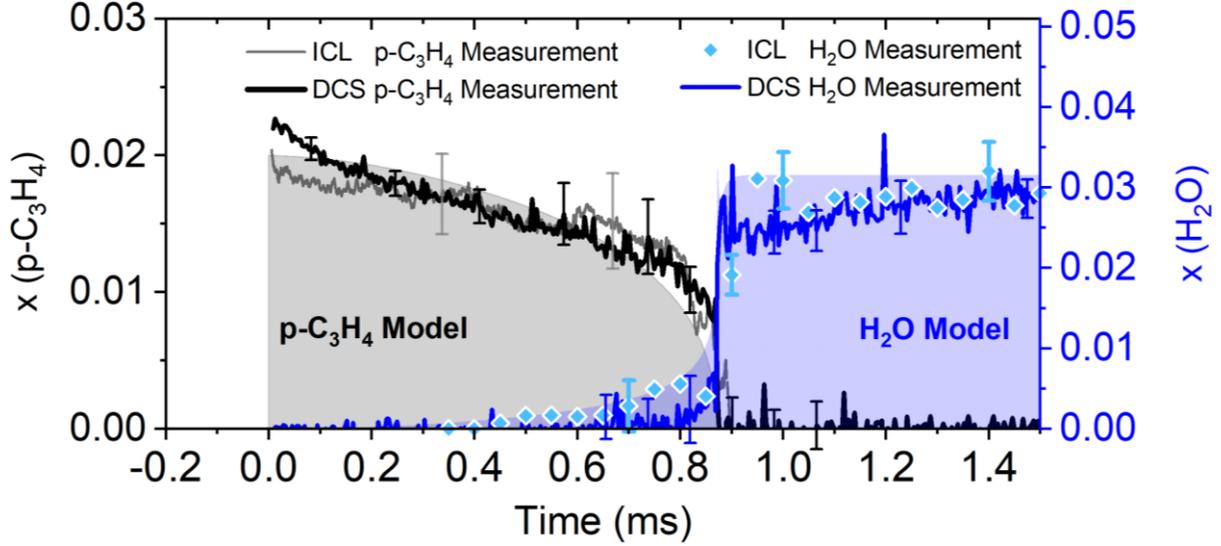

**Fig. 7.** DCS measurements of propyne oxidation with time-zero conditions of 2% p-$C_3H_4$/18% $O_2$ in Ar, 1225 K, and 2.8 atm. Speciation measurements at a 95% confidence interval and simulated results for propyne and water. All simulations used the USC Mech II kinetic mechanism and assumed constant volume [30]. Agreement at a 95% confidence interval is observed between the DCS and supporting laser measurements. While overall agreement exists between the DCS measurements and USC Mech II, key differences are visible at early times and just after ignition.

Notably, all 179 wavelengths were used in the optimization process. Each wavelength made a valuable contribution to the final mole fraction solution, however, with the specific contribution commensurate to the wavelength's SNR. By employing this weighted optimization procedure, the mole fractions of propyne and water were found to be in fair agreement with the validation diagnostics and USC Mech. II (see Fig. 7), with fluctuations in the mole fraction of water roughly correlating to the pressure shown in Fig. 6(a). In the DCS-based mole fraction measurements at early times (0 to 0.2 ms), we observe a convex decay profile that appears higher and at odds to the concave predictions of the model and ancillary propyne diagnostic. This difference after time-zero is still being investigated as to the source.

### 3.3. Uncertainty quantification

Uncertainty was propagated for the DCS and the supporting laser diagnostics to assess whether the results agree within a 95 % confidence interval. All uncertainties are included with each measurement in Fig. 7. The single-wavelength propyne diagnostic uncertainty is presented in light gray bars at 0.35 ms and 0.68 ms, the scanned-wavelength water diagnostic uncertainty as light blue bars at 0.7 ms, 1 ms, and 1.4 ms, and the DCS propyne (black) and DCS water (dark blue) uncertainty visible at all other times where ± bars are indicated.

DCS uncertainty was largely governed by the high-temperature absorption cross-section measurements used for speciation. These measurements were collected through the non-oxygenated propyne experiments and the water-generating experiments described previously. Linear temperature-dependent cross-section correlations were determined at each of the 179 wavelengths along with each correlation's uncertainty to a 95% confidence interval. Equation 2 was used to compute the cross-section ($\sigma$) uncertainty at wavelength $\lambda$ ($u_{\lambda,\sigma,95\%}$) and involved: the residual squared error for the linear fit at wavelength $\lambda$ ($RSS_\lambda$), sample number (n), and the temperature for each cross-section measurement ($T_i$). This method for uncertainty propagation was described previously in [31] and derived from [32].

$$u_{\lambda,\sigma,95\%} = t_{95\%}\left\{(RSS_\lambda)\left[\frac{1}{n} + \frac{\left(T - \frac{1}{n}\sum_{i=1}^{n}T_i\right)^2}{\sum_{i=1}^{n}T_i^2 - \frac{\left(\sum_{i=1}^{n}T_i\right)^2}{n}}\right]\right\}^{1/2} \quad \text{Eq. 2}$$

Uncertainty associated with the pressure- and path-length-normalized absorbance ($u_b$) was estimated by propagating error due to path length, pressure, and attenuated and un-attenuated laser intensities through the Beer-Lambert relation. Notably, the attenuated intensity was found to be the dominant contributor. Equation 3 was then used in the determination of the uncertainty in the mole fraction of species $i$ at time $t$ during speciation.

$$u_i(t) = \sqrt{\sum_{j=1}^{M}\sum_{k=1}^{N}\left(\frac{\partial x_i}{\partial \sigma_{j,k}}\right)^2 u_{\sigma_{j,k}}^2 + \sum_{j=1}^{M}\left(\frac{\partial x_i}{\partial b_j}\right)^2 u_{b,j}^2} \quad \text{Eq. 3}$$

The terms $\left(\frac{\partial x_i}{\partial \sigma}\right)$, $\left(\frac{\partial x_i}{\partial b}\right)$ are the derivatives of the mole fraction of species $x_i$ with respect to the cross-section correlation, $\sigma$, and pressure- and path-length-normalized absorbance, b. Notably, the uncertainties in the final measurements were calculated assuming the potential presence of methane. In these calculations, methane was added to the matrix K from Eq. 1 to assess how the measurements could be affected with a third absorber present. The propyne and water species time-histories were unaffected by this addition, however, their corresponding uncertainties were affected. The largest influence was to the propyne mole fraction uncertainty, which increased by about 5 % in the 0.2 ms proceeding ignition. No statistically significant methane was observed to form and therefore the results with no methane are shown in Fig. 7 but with the larger uncertainties included. Future studies may also consider numerically perturbing the temperature profile used in the determination of the temperature-dependent cross-sections. However, the uncertainty here was largely dominated by the uncertainty in the absorption cross-section measurements and the absorbance. Therefore, adding these additional terms will yield only minor contributions to the overall uncertainty reported.

This method of uncertainty propagation was also used for the fixed-wavelength propyne diagnostic presented as light gray in Fig. 7. This diagnostic used only one wavelength, rather than 179 for the DCS. Correspondingly, the uncertainty is much larger for the single-wavelength diagnostic due to a very large sensitivity from the term $\left(\frac{\partial x_i}{\partial b}\right)$. This observation confirms the utility of using multiple wavelengths as a means of decreasing this sensitivity and, therefore, the overall uncertainty in mole fraction measurements. In contrast, for the DCS diagnostic, improvements in the reference cross-sections will yield a largest reduction in uncertainty.

Uncertainties associated to the scanned-wavelength measurements were determined using the RSS of the Voigt fit to the absorbance trace recorded and detailed in the literature [29].

In consideration of the uncertainties presented, the DCS and supporting diagnostic measurements of both propyne and water agree within a 95 % confidence interval. However, before 0.1 ms and immediately after ignition a subtle but observable disagreement exists between the DCS and the supporting diagnostics that is still under investigation.

## 4. Time resolution and spectral accuracy

### 4.2. Time resolution

Figure 8 (a-b) present wavenumber slices of the DCS measurement at 100 µs and 4 µs time resolution, respectively, to illustrate the DCS data as a function of the averaging time. Low SNR regions between 1200-1210 cm$^{-1}$ and beyond 1230 cm$^{-1}$ exhibit high noise yet contain valuable data even at 4 µs time resolution. The blended-spectrum of the P-branch of the propyne can be seen very clearly at early times for both 100 µs and 4 µs cases. Likewise, the consumption of propyne and the final spectrum of water can be made out from the last two frames of Fig. 8 (a-b). We conclude that the 4 µs resolution data are of sufficient quality to make out the spectrum and its evolution. Although noise increases with shorter time resolution, and there are regions with very low SNR, this data demonstrates the opportunity to DCS to effectively capture reaction kinetics.

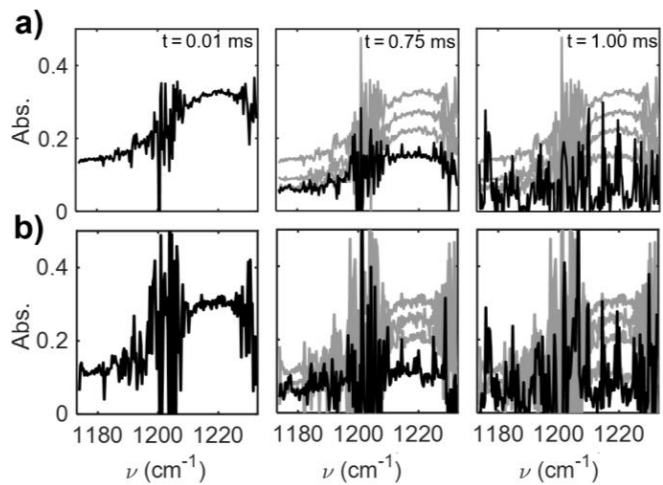

**Fig. 8.** Forward-time-averaged slices of the DCS measurements from the case study of propyne oxidation (2% p-C$_3$H$_4$/18% O$_2$ in Ar, 1225 K, and 2.8 atm) for (a) 100 µs and (b) 4 µs time-resolution. Gray traces indicate the spectrum at preceding 0.25 ms time-steps.

### 4.3. Spectral accuracy

Spectral accuracy of the spectrometer was tested both using an independent laser diagnostic for propyne and two available high-temperature spectroscopic models for water. The spectrum of high-temperature propyne showed good agreement to independent spectral measurements, confirming the spectral accuracy of the system. However, an interesting disagreement was found between the measured spectrum and available spectroscopic models for high-temperature water [27,33]. At this time, no high-temperature models exist for propyne that can be used for comparison.

*4.3.1. Independent spectral measurements / propyne*

A rapidly-tunable external-cavity QCL was used to scan over the same portion of the spectral region as the DCS. A full description of the scanned QCL laser methodology is provided in the literature [1]. Non-oxygenated mixtures of 2% p-C$_3$H$_4$/Ar were used to compare each diagnostic. Experiments using the scanned-QCL were limited to temperatures below 1130 K, where propyne was thermally stable for the 2 ms scan time. A common condition of 1120 K and 3 atm was tested for both systems and the results are shown in Fig. 9. The DCS measurement in Fig. 9 was time-averaged over the same 2 ms as the scan from the scanned-QCL. Good agreement is visible between the DCS and the QCL spectra with the only discrepancies in the two measurements corresponding to low SNR regions of the DCS. The scanned-QCL's spectral resolution is limited to 0.3 cm$^{-1}$, roughly equal to the DCS point spacing. Accordingly, the scanned-QCL was only used on the broadband spectrum of propyne and spectroscopic models used to investigate the spectrum of water.

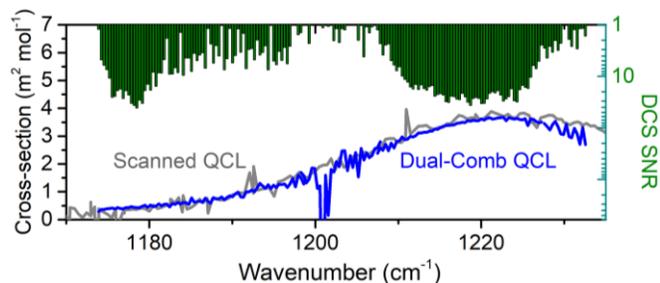

**Fig. 9.** A comparison of the high-temperature spectrum of p-C$_3$H$_4$/Ar near 1120 K and 3 atm measured using two-independent diagnostics: a scanned QCL laser (gray) [1] and the QCL DCS system (blue). SNR is shown on right axis (green) and was calculated using the DCS signal (blue) and standard deviation. Good agreement is visible with discrepancies correlated to low SNR regions.

*4.3.2. Spectral modeling / water*

Additional measurements were taken to study the spectrum of water and compare against available high-temperature spectroscopic models. Hydrogen oxidation experiments were used to generate a post-ignition condition of 4% H$_2$O, T=1700-1800 K, P = 3-4 atm, which was studied using the DCS. Hydrogen oxidation provided a controlled method to produce water that could be measured using the DCS and compared to two spectroscopic models: the HITEMP model and a recent model put forth by Polyansky et al. and part of the ExoMol database [34,35]. Publically available fortran scripts, ExoCross, were used to execute the ExoMol model in this study [36]. Conditions used for the spectroscopic simulation relied on modeled predictions of the post-ignition temperature using USC Mech II, the water mole fraction measured by the supporting water diagnostic, and pressure measured using a side-wall transducer. A resulting comparison between the DCS measurement and HITEMP modeled spectrum of high-temperature water is provided in Fig. 10. The DCS measurements use the post-ignition spectrum of hydrogen oxidation averaged from 400-600 µs. It is evident that the measured spectrum under-samples the predicted spectrum and there is poor agreement between the measured and simulated spectra. Even when considering the under-sampling rate, the difference between the modeled and measured spectrum cannot be reconciled at this time.

It is of great interest to the authors to fit high-temperature spectroscopic models to spectra measured using the DCS. This is in contrast to the method employed in the current study that involved the use a library of measured spectra and absorption cross-sections for speciation. While also very effective, an improved methodology would rely upon fitting spectroscopic models to the DCS data for reduced uncertainty in speciation and to potentially enable thermometry. However, across several species that absorb in the region (propyne, methane, and water), there is a lack of high-temperature spectroscopic data to compare to, or to found models upon, in the 1200 cm$^{-1}$ region. Many studies have begun to address this absence of data already[1,26,35,37–43]. This discrepancy suggests an opportunity for future research as measurements push to shorter timescales and higher temperatures, expanding the utility of models and DCS for use in high-temperature reaction kinetics.

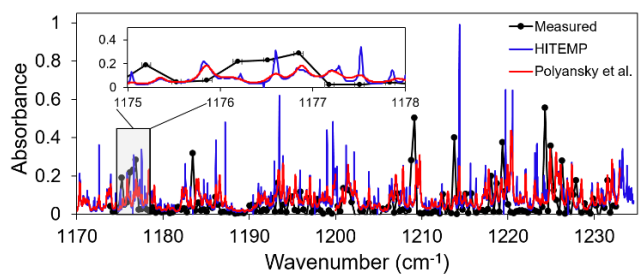

**Fig. 10.** A comparison of the measured and modeled high-temperature spectrum of 4% H$_2$O, T=1738 K, P = 3.14 atm that resulted following hydrogen oxidation, with a path length of 14.13 cm [27,33,36]. No visible agreement is observed between the measured and modeled spectra at this time. This observation identifies an opportunity to explore an understudied high-temperature band for water while also better assessing the ability for QCL-based dual combs to study very finely featured spectra.

**5. Conclusion**

A clear tradeoff exists in molecular DCS between spectral and time resolution, with the former investigated thoroughly. Despite the fewer comb teeth used in this study, the mid-infrared dual-comb system demonstrated was capable of resolving propyne oxidation kinetics at 4 µs time resolution. The time resolution and quality of the multi-heterodyne signal showed merit in the high-enthalpy test environment despite having several regions with low SNR. The DCS was validated against a suite of independent laser diagnostics with statistically significant agreement demonstrated between laser systems. The current study presented speciation of propyne and water by comparing a measured absorption surface to a library of experimentally measured, high-temperature absorption cross-sections. Agreement was also observed between the modeled and measured mole fractions of propyne and water, although slight discrepancies did exist at early times and just after ignition, a phenomenon warranting further investigation.

While agreement was obtained between this DCS and a supporting laser diagnostic to measure the broadband spectrum of propyne, a notable disagreement exists between our measured spectra of water and existing high-temperature spectroscopic models. Therefore, whether QCL-based DCS has the ability to measure the finely featured spectra of molecules such as water, remains an open question. Towards this end, future studies are needed to make additional measurements of molecules such as water or methane with finely featured spectra in this spectral region. It is also important to make use of ever-improving high-temperature spectroscopic models and cross-section databases, which are also still quite underdeveloped at these temperatures and in this region.

The case study shown here indicates that a clear opportunity exists for DCS to greatly increase the information content of the traditional laser diagnostics used in high-enthalpy and reacting test environments. As mid-infrared dual-comb spectroscopy emerges, the study of high-enthalpy reacting environments offers a wide canvas of applications on which DCS can make a lasting impact.

**Funding.** This material is based upon work supported by, or in part by, the U. S. Army Research Laboratory and the U. S. Army Research Office under contract/grant number W911NF-17-1-0420 and the Air Force Office of Scientific research, AFOSR Grant No. FA9550-16-1-0195, with Dr. Chiping Li as contract monitor.

**Acknowledgment.** We acknowledge Wey-Wey Su, Adam Susa, and Séan Cassady for their contributions to the project.

**Conflict of Interest:** No conflict of interest is reported for this work.

**References**
[1] Strand CL, Ding Y, Johnson SE, Hanson RK. Measurement of the mid-infrared absorption spectra of ethylene (C2H4) and other molecules at high temperatures and pressures. J Quant Spectrosc Radiat Transf 2019;222–223:122–9. doi:10.1016/j.jqsrt.2018.10.030.
[2] Rieker GB, Giorgetta FR, Swann WC, Kofler J, Zolot AM, Sinclair LC, et al. Frequency-comb-based remote sensing of greenhouse gases over kilometer air paths. Optica 2014;1:290. doi:10.1364/OPTICA.1.000290.
[3] Werblinski T, Fendt P, Zigan L, Will S. High-speed combustion diagnostics in a rapid compression machine by broadband supercontinuum absorption spectroscopy. Appl Opt 2017;56:4443. doi:10.1364/ao.56.004443.
[4] Schroeder PJ, Wright RJ, Coburn S, Sodergren B, Cossel KC, Droste S, et al. Dual frequency comb laser absorption spectroscopy in a 16 MW gas turbine exhaust. Proc Combust Inst 2017;36:4565–73. doi:10.1016/j.proci.2016.06.032.
[5] Coburn S, Alden CB, Wright R, Cossel K, Baumann E, Truong G-W, et al. Regional trace-gas source attribution using a field-deployed dual frequency comb spectrometer. Optica 2018;5:320. doi:10.1364/OPTICA.5.000320.
[6] Coddington I, Newbury N, Swann W. Dual-comb spectroscopy. Optica 2016;3:411–26.
[7] Picqué N, Hänsch TW. Frequency comb spectroscopy. Nat Photonics 2019;13:146–57. doi:10.1038/s41566-018-0347-5.
[8] Draper AD, Cole RK, Makowiecki AS, Mohr J, Zdanawicz A, Marchese A, et al. Broadband Dual Frequency Comb Spectroscopy in a Rapid Compression Machine. Opt Express 2018;27:10814–25.
[9] Sanders ST. Wavelength-agile fiber laser using group-velocity dispersion of pulsed super-continua and application to broadband absorption spectroscopy. Appl Phys B Lasers Opt 2002;75:799–802. doi:10.1007/s00340-002-1044-z.
[10] Klocke JL, Mangold M, Allmendinger P, Hugi A, Geiser M, Jouy P, et al. Single-shot submicrosecond mid-infrared spectroscopy on protein reactions with quantum cascade laser frequency combs. Anal Chem 2018:acs.analchem.8b02531. doi:10.1021/acs.analchem.8b02531.
[11] Shimanouchi T. Molecular Vibrational Frequencies: Consolidated Vol. I 1972:1–161.
[12] Fleisher AJ, Bjork BJ, Bui TQ, Cossel KC, Okumura M, Ye J. Mid-infrared time-resolved frequency comb spectroscopy of transient free radicals. J Phys Chem Lett 2014;5:2241–6. doi:10.1021/jz5008559.
[13] Bjork BJ, Bui TQ, Heckl OH, Changala PB, Spaun B, Heu P, et al. Direct frequency comb measurements of OD+CO-> DOCO kinetics. Science (80- ) 2016;354:444–8. doi:10.7910/DVN/BH9UXW.SUPPLEMENTARY.
[14] Ycas G, Giorgetta FR, Baumann E, Coddington I, Herman D, Diddams SA, et al. High-coherence mid-infrared dual-comb spectroscopy spanning 2.6 to 5.2 μm. Nat Photonics 2018;12:1–7. doi:10.1038/s41566-018-0114-7.
[15] Iwakuni K, Porat G, Bui TQ, Bjork BJ, Schoun SB, Heckl OH, et al. Phase-stabilized 100 mW frequency comb near 10 μm. Appl Phys B Lasers Opt 2018;124:1–7. doi:10.1007/s00340-018-6996-8.
[16] Yu M, Okawachi Y, Griffith AG, Picqué N, Lipson M, Gaeta AL. Silicon-chip-based mid-infrared dual-comb spectroscopy. Nat Commun 2018;9:6–11. doi:10.1038/s41467-018-04350-1.
[17] Muraviev A V., Smolski VO, Loparo ZE, Vodopyanov KL. Massively parallel sensing of trace molecules and their isotopologues with broadband subharmonic mid-infrared frequency combs. Nat Photonics 2018;12:209–14. doi:10.1038/s41566-018-0135-2.
[18] Villares G, Hugi A, Blaser S, Faist J. Dual-comb spectroscopy based on quantum-cascade-laser frequency combs. Nat Commun 2014;5:1–3. doi:10.1038/ncomms6192.
[19] Schliesser A, Picqué N, Hänsch TW. Mid-infrared frequency combs. Nat Photonics 2012;6:440–9. doi:10.1038/nphoton.2012.142.
[20] Hugi A, Villares G, Blaser S, Liu HC, Faist J. Mid-infrared frequency comb based on a quantum cascade laser. Nature 2012;492:229–33. doi:10.1038/nature11620.
[21] Marandi A, Leindecker NC, Pervak V, Byer RL, Vodopyanov KL. Coherence properties of a broadband femtosecond mid-IR optical parametric oscillator operating at degeneracy. Opt Express 2012;20:7255. doi:10.1364/oe.20.007255.
[22] Leindecker N, Marandi A, Byer RL, Vodopyanov KL. Broadband degenerate OPO for mid-infrared frequency comb generation. Opt Express 2011;19:6296. doi:10.1364/oe.19.006296.
[23] Villares G, Faist J. Quantum cascade laser combs: effects of modulation and dispersion. Opt Express 2015;23:1651. doi:10.1364/OE.23.001651.
[24] Faist J, Villares G, Scalari G, Rösch M, Bonzon C, Hugi A, et al. Quantum Cascade Laser Frequency Combs. Nanophotonics 2016;5:272–91. doi:10.1515/nanoph-2016-0015.
[25] Villares G, Hugi A, Blaser S, Faist J. Dual-comb spectroscopy based on quantum-cascade-laser frequency combs. Nat Commun 2014;5:1–3. doi:10.1038/ncomms6192.

[26] Es-Sebbar E, Jolly A, Benilan Y, Farooq A. Quantitative mid-infrared spectra of allene and propyne from room to high temperatures. J Mol Spectrosc 2014;305:10–6. doi:10.1016/j.jms.2014.09.004.
[27] Polyansky OL, Kyuberis AA, Zobov NF, Tennyson J, Yurchenko SN, Lodi L. ExoMol molecular line lists XXX: a complete high-accuracy line list for water. Mon Not R Astron Soc 2018;480:2597–608. doi:10.1093/mnras/sty1877.
[28] Goldenstein CS, Jeffries JB, Hanson RK. Diode laser measurements of linestrength and temperature-dependent lineshape parameters of H2O-, CO2-, and N2-perturbed H2O transitions near 2474 and 2482nm. J Quant Spectrosc Radiat Transf 2013;130:100–11. doi:10.1016/j.jqsrt.2013.06.008.
[29] Ferris AM, Streicher JW, Susa AJ, Davidson DF, Hanson RK. A comparative laser absorption and gas chromatography study of low-temperature n-heptane oxidation intermediates. Proc Combust Inst 2019;37:249–57. doi:10.1016/j.proci.2018.05.018.
[30] Wang H, You X, Joshi A. High-temperature Combustion Reaction Model of H2/CO/C1-C4 Compounds 2007. http://ignis.usc.edu/USC_Mech_II.htm (accessed April 2, 2019).
[31] Pinkowski NH, Ding Y, Johnson SE, Wang Y, Parise TC, Davidson DF, et al. A multi-wavelength speciation framework for high-temperature hydrocarbon pyrolysis. J Quant Spectrosc Radiat Transf 2019;225. doi:10.1016/j.jqsrt.2018.12.038.
[32] Coleman H, Steele G. Experimentation, validation, and uncertainty analysis for engineers. John Wiley and Sons Inc; 2009.
[33] Goldenstein CS, Miller VA, Mitchell Spearrin R, Strand CL. SpectraPlot.com: Integrated spectroscopic modeling of atomic and molecular gases. J Quant Spectrosc Radiat Transf 2017;200:249–57. doi:10.1016/j.jqsrt.2017.06.007.
[34] Rothman LS, Gordon IE, Barber RJ, Dothe H, Gamache RR, Goldman A, et al. HITEMP, the high-temperature molecular spectroscopic database. J Quant Spectrosc Radiat Transf 2010;111:2139–50. doi:10.1016/j.jqsrt.2010.05.001.
[35] Polyansky OL, Kyuberis AA, Zobov NF, Tennyson J, Yurchenko SN, Lodi L. ExoMol molecular line lists XXX: A complete high-accuracy line list for water. Mon Not R Astron Soc 2018;480:1–12. doi:10.1093/mnras/sty1877.
[36] Yurchenko SN, Al-Refaie A, Tennyson J. ExoCross: a general program for generating spectra from molecular line lists 2018. doi:10.1051/0004-6361/201732531.
[37] Ding Y, Strand CL, Hanson RK. High-temperature mid-infrared absorption spectra of methanol (CH3OH) and ethanol (C2H5OH) between 930 and 1170 cm-1. J Quant Spectrosc Radiat Transf 2018;224:396–402. doi:10.1016/j.jqsrt.2018.11.034.
[38] Almodovar CA, Su W-W, Strand CL, Sur R, Hanson RK. High-pressure, high-temperature optical cell for mid-infrared spectroscopy. J Quant Spectrosc Radiat Transf 2019. doi:10.1016/j.jqsrt.2019.04.014.
[39] Coustenis A, Boudon V, Campargue A, Georges R, Tyuterev VG. Exo-Planetary high-Temperature Hydrocarbons by Emission and Absorption Spectroscopy (the e-PYTHEAS project). Eur Planet Sci Congr 2017;11:EPSC2017-719–1, 2017.
[40] Rey M, Delahaye T, Nikitin A V., Tyuterev VG. First theoretical global line lists of ethylene ( 12 C 2 H 4 ) spectra for the temperature range 50−700 K in the far-infrared for quantification of absorption and emission in planetary atmospheres . Astron Astrophys 2016;594:A47. doi:10.1051/0004-6361/201629004.
[41] Melin ST, Sanders ST. Gas cell based on optical contacting for fundamental spectroscopy studies with initial reference absorption spectrum of H2O vapor at 1723 K and 0.0235 bar. J Quant Spectrosc Radiat Transf 2016;180:184–91. doi:10.1016/j.jqsrt.2016.04.009.
[42] Schwarm KK, Dinh HQ, Goldenstein CS, Pineda DI, Spearrin RM. High-pressure and high-temperature gas cell for absorption spectroscopy studies at wavelengths up to 8 µm. J Quant Spectrosc Radiat Transf 2019;227:145–51. doi:10.1016/j.jqsrt.2019.01.029.
[43] Alrefae M, Es-Sebbar ET, Farooq A. Absorption cross-section measurements of methane, ethane, ethylene and methanol at high temperatures. J Mol Spectrosc 2014;303:8–14. doi:10.1016/j.jms.2014.06.007.

# Appendix A

Figure A.1 presents the wavenumber- and time-dependent SNR of the system during the propyne oxidization reaction. These SNR values were used in the formulation of the diagonal weighting matrix used in the speciation calculations.

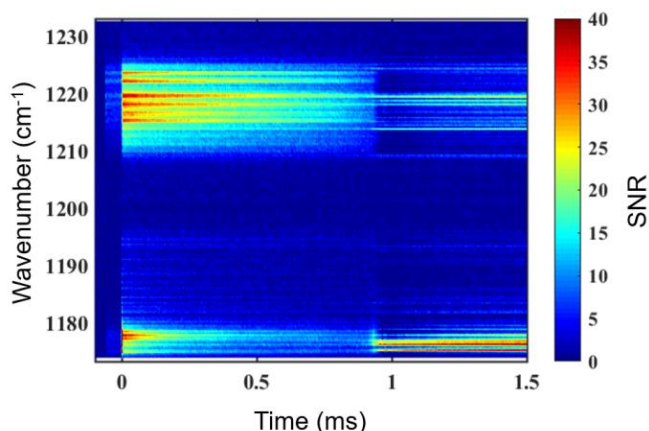

**Fig. A.1.** SNR from a single-shot measurement of propyne oxidation with time-zero conditions of 2% $pC_3H_4$/18% $O_2$ in Ar, 1225 K, and 2.8 atm. SNR is shown using the color bar on the right.